\documentclass[aps,pre,twocolumn,floatfix]{revtex4}

\usepackage{graphicx}
\usepackage{dcolumn}
\usepackage{bm}

\newcommand{\pdiffl}[2]{\frac{\partial #1}{\partial #2}}

\newcommand{\dfrac}[2]{\displaystyle\frac{#1}{#2}}

\begin{document}


\title{Non-Iterative Characteristics Analysis \\ for Ramp Loading with a Window}

\date{November 6, 2018, revisions to February 12, 2019 -- LLNL-JRNL-767557}

\author{Damian C. Swift}
\email{dswift@llnl.gov}
\author{Dayne E. Fratanduono}
\author{Evan A. Dowling\footnote{%
Current affiliation: University of Maryland -- College Park, Maryland, U.S.A.
}}
\author{Richard G. Kraus}
\affiliation{%
   Physics Division, Lawrence Livermore National Laboratory,
   7000 East Avenue, Livermore, California 94551, U.S.A.
}

\begin{abstract}
Ramp compression experiment are used to deduce the relation between
compression and normal stress in a material,
by measuring how a compression wave evolves as it propagates through
different thicknesses of the sample material.
The compression wave is generally measured by Doppler velocimetry
from a surface that can be observed with optical or near-optical photons.
For high-pressure ramp loading, the reflectivity of a free surface
often decreases as it is accelerated by the ramp wave,
and window materials transparent to the probing photons 
are used to keep the surface flatter and preserve its reflectivity.
We previously described a method of analyzing ramp-wave data measured at
the free surface which did not require numerical iteration.
However, this method breaks down when the pressure at the surface changes
and hence cannot be used for data taken with a finite-impedance window.
We have now generalized this non-iterative analysis method to apply to
measurements taken through a window.
Free surfaces can be treated seamlessly,
and the need for sampling at uniform intervals of velocity has been removed.
These calculations require interpolation of partially-released states
using the partially-constructed stress-compression relation, 
making them slower than the previous free-surface scheme, 
but they are still much more robust and fast than iterative analysis.
\end{abstract}


\maketitle

\section{Introduction}
Ramp loading, or shockless compression, 
is used increasingly in equation of state (EOS) studies 
to measure the compressibility of matter
along a continuous path that typically falls close to an isentrope
\cite{Hall2001}.
Notable aspects of ramp loading experiments are that the sample is heated
less than in shock loading, and a range of compressions are probed in a
single experiment.
Several methods have been used to apply ramp loading to a sample,
including chemical explosives \cite{Barnes1974},
pulsed magnetic fields \cite{Hall2001},
and laser pulses converted to pressure in a variety of ways
\cite{Edwards2004,Swift_lice_2005,Smith2007,Bradley2009}.

Ramp loading does not generally induce perfectly
isentropic compression in the sample because of irreversible processes 
such as plastic flow, although the ramp-loaded states can be
corrected to find the isentrope if the constitutive response of the
sample is known \cite{Kraus2016}.
For convenience, we  will use `isentrope' to refer below to states obtained
by ramp loading.
Similarly, although the component of stress normal to the loading direction
is the relevant value, and in general differs from the pressure because
of material strength effects, for convenience we refer to normal stress
as pressure.

The ramp wave is almost invariably analyzed from measurements of the 
velocity history at different distances into the sample,
using interferometry of near-optical laser probes \cite{Barker1974,Strand2006}.
Most materials are or become optically opaque,
so the measurement must be performed at a surface bounded by
a transparent window or a vacuum.
Generally, this interface introduces an impedance mismatch that complicates the
measurement by reflecting waves back into the sample.
The impedance mismatch must be taken into account when inferring the isentrope
of the sample, and the inferred isentrope therefore depends on knowledge of the
isentrope of the window.
In this respect, a vacuum is preferable as its mechanical properties are 
perfectly defined.
However, in practice it is found that the optical reflectivity of the
free surface of the sample is often degraded, particularly at high pressures.
Possible causes include
changes in the inherent reflectivity of the material
for instance by changes in conductivity caused by phase transitions,
and by disruption of the surface as by ejecta from microscopic relief
features such as machining grooves.
In any case, transparent windows are employed widely as a palliative to
confine the surface and maintain usable signal levels.

Ramp-loading data have generally been analyzed using
an iterative technique in which the isentrope is assumed, and corrections
are made repeatedly using the ramp data until the
modified isentrope converges \cite{Rothman2006}.
With such iterative refinement techniques, there is always a potential
concern that the solution found may depend on the isentrope initially assumed,
or on the parameters of the iterative scheme,
or that the algorithm may fail to converge.
We previously described a recursive, non-iterative analysis method
\cite{Swift2018}
that gives the same result with greater stability and less computational effort,
but was limited to free surface measurements sampled at uniform intervals
of velocity.
Here we generalize the previous method to windowed measurements;
the generalized method of analysis is also valid for non-uniform sampling
in velocity, which aligns more naturally with experimental data usually taken
with uniform sampling in time.

The research described here was presented at the 2017 conference
of the American Physical Society Topical Group on Shock Compression of
Condensed Matter \cite{DowlingSCCM2017}.
More detailed comparisons with hydrocode simulations were made subsequently,
and are included below.

\section{Isentrope measurement from ramp loading}
\begin{figure}
\begin{center}\includegraphics[scale=0.72]{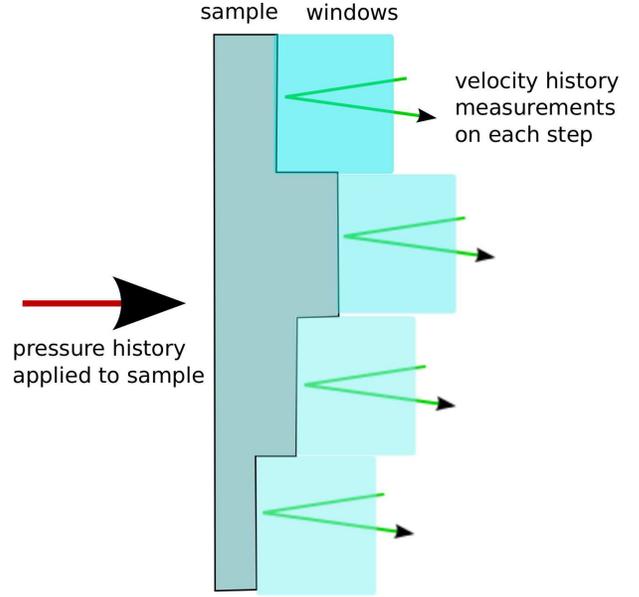}\end{center}
\caption{Schematic of ramp wave experiment with window, used to deduce EOS.}
\label{fig:rampschem}
\end{figure}

The stress-density relation for a material from a given initial state
can be deduced from the evolution of a ramp load as it propagates through
a sample of the material.
The longitudinal sound speed generally varies with compression,
so the evolution of a ramp in bulk matter is related directly to the sound speed
and hence the normal stress.
In a typical experimental configuration, the particle velocity
history $u(t)$ is measured at each of a set of steps of different thickness $x$
(Fig.~\ref{fig:rampschem}).
If the window material through which each step is observed were a perfect
impedance match to the sample, the particle speed measured at the interface
at a given pressure would be the same as in the bulk material, and the sound
speed $c$ could be found from the time at which the particle speed $u_p$
appeared at steps of different thickness,
\begin{equation}
c(u_p)=\Delta X/\Delta t(u_p;X),
\end{equation}
However, usually the window is not a perfect impedance match,
and forward-propagating characteristics from the ramp reflect back from the
interface, perturbing the ramp and hence the velocity history measured
at each step.
The next section describes our new algorithm for correcting for
the characteristics reflected from the impedance mismatch from the window.

As defined here,
$c$ is the longitudinal Lagrangian sound speed, 
i.e. with respect to uncompressed material.
In the absence of elastic contributions,
$c$ is related to the sound speed calculated from
the EOS, $c_e=c\rho_0/\rho$ where $\rho$ is the mass density
and $\rho_0$ its initial value.
The normal stress $p$ (or pressure, again if elasticity is neglected)
can be calculated from $c_e$ from the Riemann integral
\begin{equation}
p=\int \rho c_e\,du_p; \quad \rho=\int \frac{\rho}{c_e}du_p
\end{equation}
where the EOS sound speed is
\begin{equation}
c_e\equiv\sqrt{B_s/\rho}
\end{equation}
and $B_s$ is the isentropic bulk modulus,
$\left.\rho\,\partial p/\partial\rho\right|_s$.

In the iterative analysis used previously \cite{Rothman2006},
an estimate of the isentrope is used to correct for the reflected
characteristics and hence obtain a new estimate of the isentrope.
This process is repeated until a converged result is obtained.
We showed previously \cite{Swift2018} that, in the absence of a window,
the correction for perturbing characteristics can be made in a non-iterative,
deterministic way.
Here we extend that analysis to the modified characteristics induced by 
a window.

\section{Ramp characteristics at an interface with a window}
For experiments with a window to enable the isentrope of the sample
to be deduced, the window's EOS must be known,
along with its refractive index as a function of compression
along its isentrope.
The refractive index is often represented as a correction function 
that maps apparent particle speed (as if the refractive index were unity,
as for a free surface) to the actual particle speed;
physically, this correction involves a combination of the refractive index
and the EOS \cite{Dolan2006}.
At the interface,
any given incoming characteristic has been affected by the reflection
of all previous characteristics,
but not by the reflection from any subsequent ones.
If the window has a lower impedance than the sample,
the reflected characteristics have a lower pressure than the subsequent
incident characteristics,
so the correction can be made deterministically using the corrected speed
at the appropriate pressure that has already been explored.
If the window has a higher impedance,
the reflected characteristics have a higher pressure than the incident ones,
so the correction must be made either by assuming the isentrope and iterating,
or assuming a form of extrapolation from the previously-reconstructed
region, which may be possible to adequate accuracy without iteration.
From this point, the effect can be corrected for recursively, similarly to
the free surface case \cite{Swift2018}.

\begin{figure}
\begin{center}\includegraphics[scale=0.80]{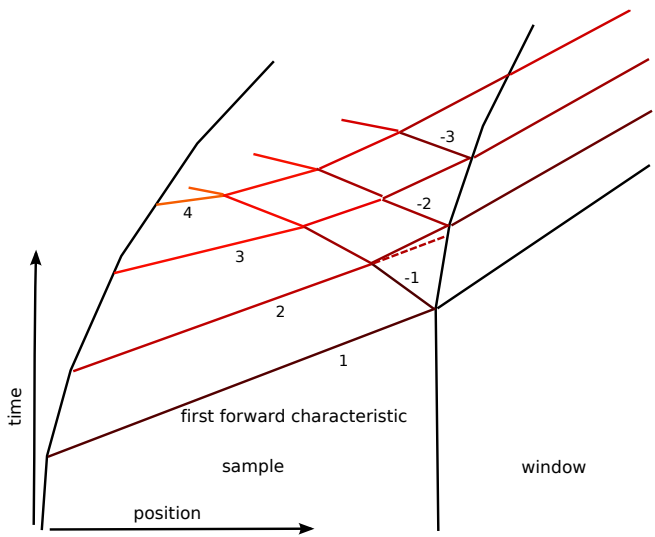}\end{center}
\caption{Interaction between successive reflected characteristics.
   Successive forward-moving loading characteristics are numbered in
   sequence; negative numbers are the corresponding reflected characteristics.
   The dotted line represents the trajectory of characteristic 2 if it had not
   been perturbed by the reflection of characteristic 1.}
\label{fig:characs}
\end{figure}

The recursive correction proceeds by considering pairs of (position,time)
points corresponding to successive forward-going characteristics $i$ and $i+1$ 
either
at the interface or after correcting both for perturbations from the same
number ($j-1$) of backward-going reflective characteristics, and deducing the
intersection point between later characteristic
and the next ($j$th) backward-going characteristic.
In Lagrangian coordinates, i.e. the position of each element of the sample
before loading starts, the next intersection is at
\begin{eqnarray}
t_{i+1,j,s} & = & \frac{x_{i,j,s}-x_{i+1,j-1,s}}{2 c_i}
   +\frac{t_{i,j,s}+t_{i+1,j-1,s}}2 \\
\label{eq:intt}
x_{i+1,j,s} & = & x_{i+1,j-1,s} - c_i (t_{i+1,j,s} - t_{i+1,j-1,s})
\label{eq:intx}
\end{eqnarray}
where the subscript $s$ means that these equations are applied
separately to data from each step.
\cite{Swift2018}.
This correction can be made for each step, and the corresponding correction
can be made for all subsequent characteristics
(Figs~\ref{fig:characs}).

Unlike the free surface case,
the corrections for all intersecting left-going characteristics
are applied to each right-going characteristic,
and the corresponding intersection points stored,
before the next right-going characteristic is processed.
In contrast, for a free surface, every right-going characteristic can be
processed to remove the effect of each left-going characteristics in turn,
starting with the reflection from the surface and moving successively left
\cite{Swift2018};
there is no need to store data at the intersection of left-going characteristics
once the correction for each has been made.
The requirement to store more intermediate data is not a practical limitation
for current or likely experiments, but it does make the software implementation
more complicated.

As in the free surface case,
after each characteristic has been corrected for the
reflections from all preceding characteristics, its Lagrangian wave speed
(again meaning with respect to uncompressed material) 
can be calculated from the corrected position-time data
for two or more steps $s$:
\begin{equation}
c_i = \dfrac{\partial_s x_{i,i-1,s}}{\partial_s t_{i,i-1,s}}
\end{equation}
where the partial derivative is used to represent linear fitting over the
time -- position data,
or
\begin{equation}
c_i = \dfrac{x_{i,i-1,2}-x_{i,i-1,1}}{t_{i,i-1,2}-t_{i,i-1,1}}
\end{equation}
for two steps.

\begin{figure}
\begin{center}\includegraphics[scale=0.72]{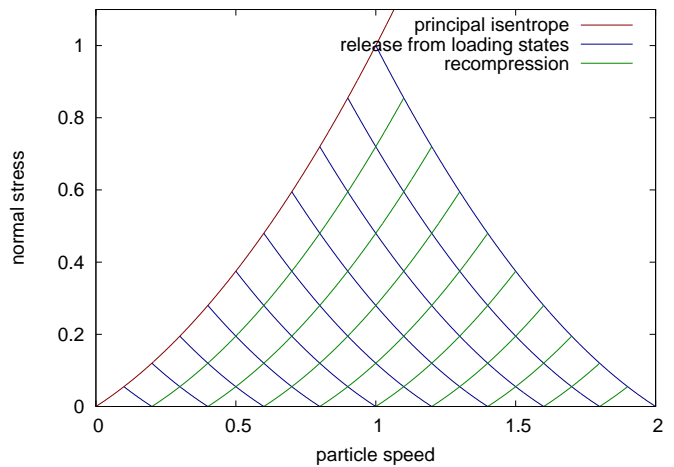}\end{center}
\caption{Pressure-particle speed states occurring with free-surface velocity
   sampled at uniformly-spaced intervals.}
\label{fig:puequal}
\end{figure}

\begin{figure}
\begin{center}\includegraphics[scale=0.72]{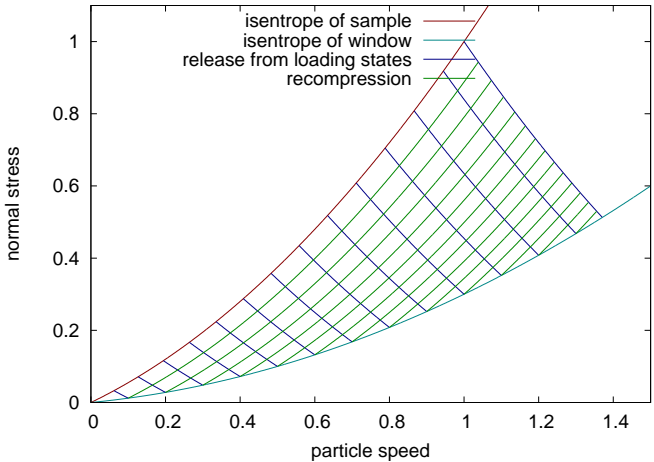}\end{center}
\caption{Pressure-particle speed states occurring with velocity
   measured at interface with window.}
\label{fig:puwin}
\end{figure}

In the free-surface analysis \cite{Swift2018}, 
we found that the algorithm required the free surface velocity history to be
determined at equal increments of velocity,
in common with preceding iterative algorithms \cite{Rothman2006}.
This limitation may be inconvenient in practice, 
because experimental diagnostics of
velocity typically determine the history at (at least approximately)
equal increments of time.
In a free-surface experiment, any characteristic reaching the surface travels
at the zero-pressure longitudinal wave speed.
With equal increments of velocity, 
the speed of characteristics further from the surface depends only on the
number of interactions with lower-pressure characteristics,
and thus the intersections between characteristics occur at the same set
of velocities as are considered at the free surface, and the same set
of pressures as are inferred along the isentrope
\cite{Swift2018} (Fig.~\ref{fig:puequal}).
In the presence of a window,
each reflected characteristic starts from the isentrope of the window
at the pressure corresponding to the measured interface velocity.
Thus the symmetry between right- and left-moving characteristics is broken,
and neither the velocities nor pressures of intersections between
characteristics fall into a set defined by its boundary values
(Fig.~\ref{fig:puwin}).

\begin{figure}
\begin{center}\includegraphics[scale=0.72]{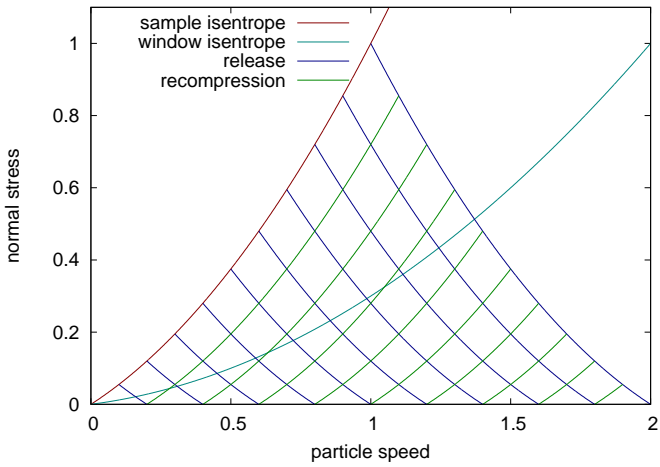}\end{center}
\caption{Pressure-particle speed states along window isentrope,
   corrected so that effective free-surface velocity
   would be sampled at uniformly-spaced intervals.}
\label{fig:puequalwin}
\end{figure}

In principle, the window case might be approached by interpolating
velocities at the interface to find times corresponding to evenly-spaced
velocities if it were a free surface.
However, a complication is that forward- and backward-going characteristics
would intersect the window isentrope at different times.
At any point in the analysis,
the fictitious sections of isentrope below the window isentrope
would be translations of sections of the isentrope already recovered
from the data, so this process would still be deterministic.
(Fig.~\ref{fig:puequalwin}.)

\begin{figure}
\begin{center}\includegraphics[scale=0.60]{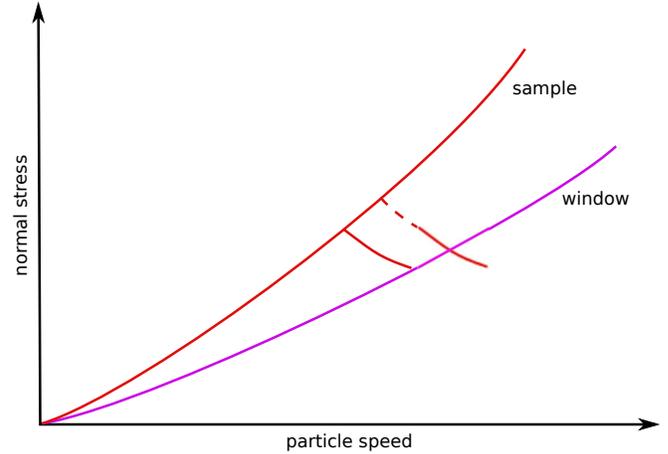}\end{center}
\caption{Reflection and translation of partially-reconstructed isentrope
   to deduce next state along isentrope.}
\label{fig:purefl}
\end{figure}

However, we chose instead to develop a method to solve the problem of
the skewed $p-u$ mesh (Fig.~\ref{fig:puwin}) as this would also apply to
the case of free-surface measurements at non-uniform intervals of velocity.
Naively, one might think that, during the reconstruction of the isentrope
of the sample,
and even with a window of lower impedance than the sample,
calculating the intersection from the previous point on the sample's isentrope 
and the last point on the backward-propagating characteristic
involves extrapolating to the next higher pressure.
However, the solution at the next point can be found by translating the
release isentrope
-- the reflection of the partly-constructed isentrope -- 
through the next state at the interface,
and considering a symmetric compression-release cycle to determine the
next state on the isentrope of the sample (Fig.~\ref{fig:purefl}).
Translation is described by
\begin{equation}
\pdiffl{u_{\mbox{trans}}}{u_{\mbox{win}}}=1+\left(\pdiffl pu\right)_{\mbox{win}}
   \left(\pdiffl pu\right)^{-1}_{\mbox{sam}}
\end{equation}
evaluated at the pressure along the isentrope of the window at $u_{\mbox{win}}$.
Then, the next state on the isentrope of the sample is at particle velocity
increased by $\Delta u=\Delta u_{\mbox{trans}}/2$ or
\begin{equation}
u=\dfrac{u_{\mbox{win}}+u_{\mbox{sam}}(p_{\mbox{win}})}2.
\end{equation}
This procedure exactly reproduces the limiting cases of a free surface 
($p_{\mbox{win}}=0 \quad\forall u_{\mbox{win}}$) and of an impedance-matched 
window.

As with our previous analysis,
the step height only enters as a difference, so absolute step heights are
irrelevant so long as the relative heights are known.

For the first characteristic, the analysis was performed completely as
described above, and also with the zero-pressure value of longitudinal
wave speed $c_0$ supplied rather than calculated.
This modification is appropriate for experimental data in which 
the uncertainty in $c_0$ derived from low-pressure measurements may be
significantly less than the value inferred from ramp data,
which is some average of the values at zero pressure and the pressure
of the first measurable acceleration.

\section{Calculation of compression and normal stress}
The calculation of compression and normal stress is similar to the
free surface case \cite{Swift2018}, 
except that the particle speed $u_{\mbox{sam}}$ of states along the
isentrope of the sample is obtained from the analysis above and not by
dividing the free surface velocity by two.
As before, the mass density changes with particle speed as
\begin{equation}
\pdiffl\rho{u_p} = \frac\rho{c_e}
\end{equation}
where the instantaneous longitudinal wave speed $c_e=c \rho_0/\rho$.
The mass density can be integrated implicitly through the characteristics,
\begin{equation}
\rho_{i+1} = \rho_i \frac{1+\Delta u_p/2 \bar c_e}{1-\Delta u_p/2 \bar c_e}
\end{equation}
where $\Delta u_p$ is the difference in inferred particle velocity between characteristics $i$ and $i+1$,
and $\bar c_e$ is the average longitudinal wave speed.

The corresponding rate of change of pressure through the characteristics is
\begin{equation}
\pdiffl p{u_p} = \rho c_e,
\end{equation}
which can be integrated numerically as
\begin{equation}
p_{i+1} = p_i + \Delta u_p \bar\rho \bar c_e.
\end{equation}

\section{Analysis of simulated data}\label{sec:test}
\begin{figure}
\begin{center}\includegraphics[scale=0.72]{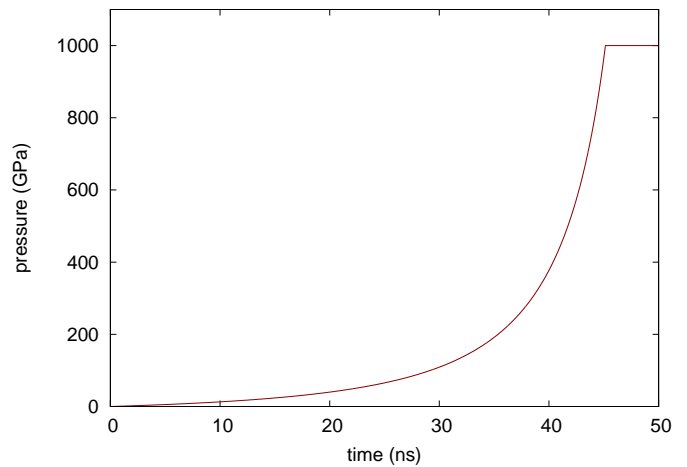}\end{center}
\caption{Loading history applied to the Cu to generate simulated data.}
\label{fig:load}
\end{figure}

\begin{figure}
\begin{center}\includegraphics[scale=0.72]{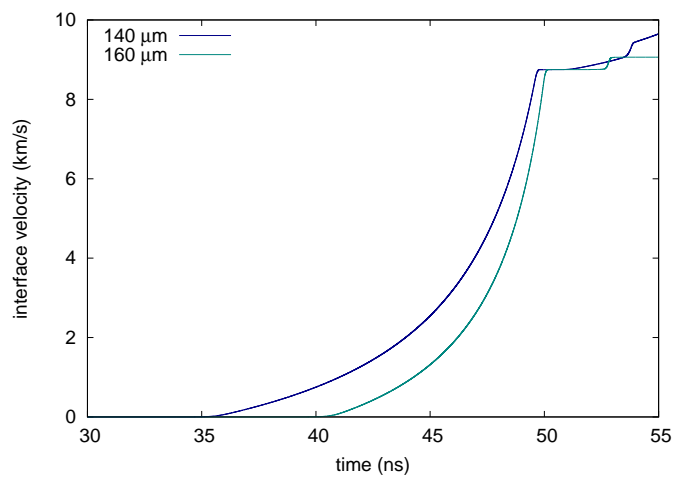}\end{center}
\caption{Simulated velocity histories at the Cu/LiF interface
   for steps of different thickness, showing steepening of ramp wave.}
\label{fig:velhist}
\end{figure}

As a test case, we consider the analysis of simulated data for Cu
ramp-loaded to 1\,TPa, with a LiF window.
The Cu and LiF were treated as behaving according to analytic EOS of the 
Gr\"uneisen form \cite{Steinberg1996}.
The loading history was chosen, as in the test case for the free surface
analysis \cite{Swift2018},
to be the ideal shape for all the characteristics to cross
to form a shock at the same point in the material \cite{Swift_idealramp_2008},
if unperturbed by the effect of the window.
The ideal shape was scaled so that the shock formation distance was
200\,$\mu$m (Fig.~\ref{fig:load}),
with suitable steps being 140 and 160\,$\mu$m.
Simulated experimental data were generated from continuum mechanics simulations 
of the velocity history at the surface of each step (Fig.~\ref{fig:velhist}).
The simulations were performed using a Lagrangian hydrocode
with a second-order predictor-corrector numerical scheme
using artificial viscosity to stabilize the flow against unphysical 
oscillations \cite{LAGJ}.
The simulations used a spatial resolution of 0.5\,$\mu$m.
Gaussian noise was added to the data in time or velocity.

\begin{figure}
\begin{center}\includegraphics[scale=0.72]{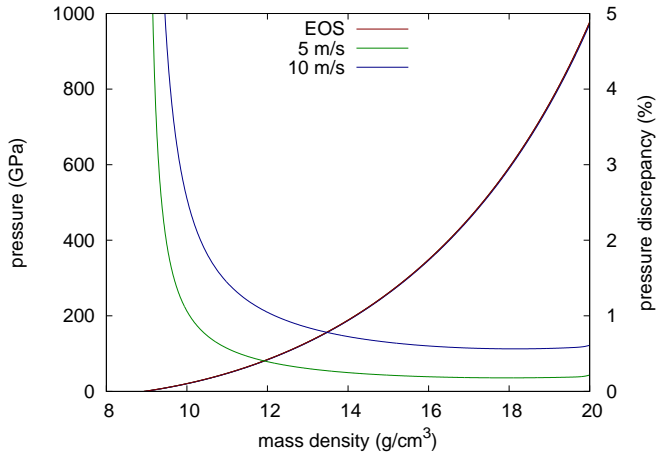}\end{center}
\caption{Isentrope obtained by direct integration and by recursive 
   characteristics analysis, with simulated data sampled at different
   intervals of velocity.}
\label{fig:analysis}
\end{figure}

The recursive analysis method was 
applied to the simulated data, produced at different resolutions,
and with or without noise.
At low pressures, the absolute discrepancy was small but the fractional
discrepancy large because of the inaccuracy in determining the precise time
at which acceleration stated.
As the pressure along the isentrope rose, the fractional discrepancy 
fell rapidly to an asymptote of a fraction of a percent.
The discrepancy was reduced when the simulated data were generated with
finer spatial resolution, implying that part of the discrepancy was
the finite precision of the hydrocode's numerical scheme.
The discrepancy was also smaller when the interface velocity history
was sampled at smaller intervals of time or velocity.
(Fig.~\ref{fig:analysis}.)

We also tested the operation of the analysis algorithm on simulated data
for a free surface, tabulated at non-uniform intervals of velocity.
The algorithm reproduced the result obtained with the previous
algorithm requiring uniform velocity intervals \cite{Swift2018}. 

Compared with the recursive free surface analysis,
the greater complexity of the window case increased the execution time
significantly.
For a typical experimental dataset, with the interface velocity sampled 
at a several hundred instants of time,
the window algorithm took around 100\,ms, compared with $\sim$10\,ms
for the free surface algorithm.
However, this is still several thousand times faster than the typical
execution time of iterative schemes.
For a high resolution dataset, with several thousand points from each step,
the increase in speed would be proportionately larger.

\section{Conclusions}
We have developed a non-iterative algorithm -- deterministic, and based
on recursion over characteristics -- for analyzing ramp-loading data,
which gives the same results as the iterative algorithm generally used.
No numerical problems were found in processing data with noise,
though the data had to be filtered to be monotonic.

Comparing with simulated data representative of real experiments,
the recursive algorithm was several orders of magnitude faster than
the iterative algorithm, and performed more stably, giving a solution
in cases where the iterative algorithm failed to converge.
Unlike an iterative solution, the recursive algorithm does not require
an estimate of the solution to be made as a starting condition, which
is undesirable as it may bias the solution.

Even with a window,
the deduction of a stress-density relation for a material from surface
velocity measurements is inherently deterministic.

\section{Acknowledgments}
Jean-Paul Davis prompted this study by asking
whether recursion works with a window.
We appreciate the support of
Tom Arsenlis, Jim McNaney, and Dennis McNabb, for encouragement and funding
via the National Ignition Facility High-Z campaign.
This work was performed under the auspices of the
U.S. Department of Energy by Lawrence Livermore National Laboratory
under Contract DE-AC52-07NA27344.

\end{document}